\documentclass[twocolumn,apj,tighten]{emulateapj-rtx4}
\usepackage{amssymb}

\usepackage{graphicx}
\usepackage{latexsym}
\usepackage{mathrsfs}
\usepackage{calrsfs}
\usepackage{amsfonts}

{\count255=\time\divide\count255 by 60 \xdef\hourmin{\number\count255}
  \multiply\count255 by-60\advance\count255 by\time
  \xdef\hourmin{\hourmin:\ifnum\count255<10 0\fi\the\count255}}

\newcommand{\nn}{\nonumber \\ }

\def\rd{ {\rm d}}
\def\vev#1{ \left\langle #1 \right \rangle }
\def\abs#1{ \left| #1 \right| }

\submitted{}

\begin{document}

\title{The Topology and Size of the Universe from the Cosmic Microwave Background}

\author{Grigor Aslanyan}
\author{Aneesh V.~Manohar}

\affiliation{Department of Physics, University of California at San Diego,
  La Jolla, CA 92093\vspace{4pt} }
  

\begin{abstract}
We study the possibility that the universe has compact topologies $\mathbb{T}^3$, $\mathbb{T}^2\times\mathbb{R}^1$, or $S^1\times\mathbb{R}^2$ using the seven-year WMAP data. The maximum likelihood 95\% confidence intervals for the size $L$ of the compact direction are $1.7 \le L/L_0 \le 2.1$, $1.8  \le L/L_0 \le  2.0$, $1.2  \le L/L_0 \le 2.1$ for the three cases, respectively, where $L_0=14.4$\,Gpc is the distance to the last scattering surface. An infinite universe is compatible with the data at $4.3\,\sigma$. We find using a Bayesian analysis that the most probable universe has topology $\mathbb{T}^2\times\mathbb{R}^1$, with $L/L_0=1.9$.
\end{abstract}


\maketitle

\section{Introduction}

General relativity is a local theory and does not predict or constrain the global properties of the spacetime manifold describing our universe, which  have to be constrained through observations. Considerations of non-trivial spatial topology go back to as early as 1900 by Schwarzschild~\citep{schwartzschild:1900} (English translation~\citep{1998CQGra..15.2539S}). In 1917 de~Sitter noted that Einstein's equations of general relativity did not constrain the global structure of the spacetime~\citep{deSitter:1917}, while in 1924, Friedmann realized that non-positive curvature spaces could also have compact topology~\citep{springerlink:10.1007/BF01328280}. A number of other papers were published in the following years (for reviews on cosmic topology see, e.g.~\citep{LachiezeRey:1995kj,Luminet:2008ew,Starkman:1998qx,Reboucas:2004dv,Roukema:2000jq,Blanloeil:2000jp,Levin:2001fg,Barrow:2001hu}, and references therein) but the subject remained purely speculative until very recent accurate measurements of the CMB (cosmic microwave background) anisotropies by the COBE~\citep{Bennett:1996ce} and WMAP~\citep{Jarosik:2010iu,Larson:2010gs,Komatsu:2010fb} satellites, and other observational advances in astrophysics and cosmology.

Various models for the topology of the universe (for the classification of different possible topologies see, e.g.~\citep{Barrow:2001hu}) have been extensively studied recently and compared to the experimental data. The two most important ones are the Poincar\'{e} dodecahedral space and the $3$-torus $\mathbb{T}^3$; these models are in best agreement with the experimental data. The Poincar\'{e} dodecahedral space arises by slicing the $3$-sphere $S^3$ and thus has positive curvature, while the $3$-torus is obtained by slicing infinite Euclidean space $\mathbb{R}^3$ and therefore is flat. Theoretical arguments about quantum creation of the universe favor the flat case. Based on the Wheeler-DeWitt equation, Linde has argued~\citep{Linde:2004nz} (see also~\citep{zeldovich:1984}) that compact flat universes are much easier to create than other models, and can naturally provide initial conditions for the onset of inflation. Furthermore, Roukema constructed a measure on the set of compact manifolds and showed that non-flat models almost never occur while flat models occur almost certainly~\citep{Roukema:2009rd} (see also~\citep{Roukema:2010mw} for a discussion of the Poincar\'{e} dodecahedral space versus the $3$-torus). Following these arguments, we will focus our attention entirely on the flat case in this work. Details on the Poincar\'{e} dodecahedral space and experimental data analysis are given in~\citep{Luminet:2003dx,Roukema:2009sg,Roukema:2004iu,Roukema:2008pm,Caillerie:2007gd,Lew:2008yz,Aurich:2004fq}, and other topologies in~\citep{Aurich:2010wf,Niarchou:2007nn,Aurich:2005hg,Aurich:2005ij,Cresswell:2005sh,Aurich:2004ke,Roukema:2004qh,Aurich:2004xa,Riazuelo:2003ud,Gomero:2002ir,Bond:1999te,Mota:2005wc,Gundermann:2005hz}.

We analyze three different flat topologies in this work, $\mathcal{M}_0=\mathbb{T}^3$, $\mathcal{M}_1=\mathbb{T}^2\times\mathbb{R}^1$, and $\mathcal{M}_2=S^1\times\mathbb{R}^2$, where the subscript denotes the number of non-compact directions. We will generically refer to all three cases as a torus. Usual flat space is $\mathcal{M}_\infty=\mathbb{R}^3$. The topology of the $3$-torus $\mathbb{T}^3$ can be obtained by identifying the opposite edges of a parallelepiped. We only consider the simplest case of a rectangular parallelepiped with equal side lengths $L$. This has the highest number of symmetries which helps reduce the computational time. Moreover, it has been argued in \citep{Weeks:2003xq} that only in well-proportioned spaces is the quadrupole of the CMB temperature-temperature correlation function suppressed compared to the infinite universe. The surprisingly low observed quadrupole is one of the motivations to invoke a compact topology. For our case of the $3$-torus, well-proportioned means that all three sides should be approximately equal. The topologies of the spaces $\mathbb{T}^2\times\mathbb{R}^1$ and $S^1\times\mathbb{R}^2$ are obtained by compactifying only $2$ or $1$ dimension respectively. Again, following the argument of \citep{Weeks:2003xq} and for the sake of simplicity, we consider only the case where the compactified dimensions of $\mathbb{T}^2\times\mathbb{R}^1$ have the same size. The size of the compact directions will be denoted by $L$. As $L \to \infty$ all three manifolds reduce to infinite flat space $\mathbb{R}^3$.

Different approaches have been proposed for extracting information about the topology of the universe from the experimental data, the two most important ones being the circles-in-the-sky test and the analysis of the CMB power spectrum. The basic idea of the circles-in-the sky test is that if the global structure of the space is smaller than the distance to the LSS (last scattering surface), then the LSS will self-intersect in circles, producing correlations between circles with different centers. The detection of such circles can reveal the global properties of space (for detailed description of the method see, e.g.~\citep{Cornish:1997ab,Mota:2010jb,Mota:2008jr,PhysRevD.70.083001,Levin:2004wt}).  The main disadvantage of the method is that it cannot be used if the size of the universe is bigger than the observable part of it (the distance to LSS). The one-year WMAP data has been analyzed with this method for signatures of non-trivial spatial topology \citep{Cornish:2003db}, ruling out the possibility of compact spaces with a length-scale smaller than $24$\,Gpc. This limit has been extended by about $10\%$ by the authors of \citep{ShapiroKey:2006hm} who have also ruled out the possibility of Poincar\'{e} dodecahedral space. The authors of \citep{2011MNRAS.tmp..137B} have analyzed the most recent seven-year WMAP data with this method, putting a lower bound of about $27.9$\,Gpc on the size of the fundamental domain for a flat universe. 

The low-$l$ (i.e.\ large scale) portion of CMB correlations is sensitive to the topology of space, which gives rise to another method for detecting the topology. The torus preserves the homogeneity of infinite space but breaks rotational invariance. This implies that the power spectrum of CMB temperature-temperature correlations does not contain all the possible information since the off-diagonal elements of the covariance matrix in the spherical harmonics expansion are non-zero in general, while the diagonal elements with equal $l$ and different $m$ values are not all equal to each other (see section \ref{covariance_matrix} for more details). Moreover, in \citep{Phillips:2004nc} it has been argued that the off-diagonal elements contain more information than the diagonal ones if the side length of the torus is less than twice the distance to the last scattering surface. Therefore, to gain all the possible information from the correlations of CMB anisotropies, one has to consider the full covariance matrix rather than just the power spectrum.

The CMB correlation functions have been previously used to analyze COBE~\citep{deOlivieraCosta:1994eb,deOliveiraCosta:1995td}, one-year WMAP~\citep{Phillips:2004nc,Kunz:2005wh} and three-year WMAP~\citep{Aurich:2007yx} data. The lower bound on the side length $L$ of $\mathbb{T}^3$ obtained from COBE data~\citep{deOlivieraCosta:1994eb} is $L>4.32h^{-1}$\,Gpc at $95\%$ confidence, and $L>5.88h^{-1}$\,Gpc at $68\%$ confidence. For $\mathbb{T}^2\times\mathbb{R}^1$ and $S^1\times\mathbb{R}^2$ the lower bound obtained from COBE~\citep{deOliveiraCosta:1995td} is $L>3.0h^{-1}$\,Gpc at $95\%$ confidence. The authors of~\citep{Phillips:2004nc} have obtained higher bounds for $\mathbb{T}^3$; $L > 1.2\, L_0$ at $95\%$ confidence and $L>2.1\,L_0$ at $68\%$ confidence.\footnote{We will give lengths in terms of $L_0=14.4$\,Gpc, the distance to the last scattering surface. The Hubble length is $H_0^{-1}= 2.998/h_0=4.266\, (0.703/h_0)$\,Gpc.} They have also found that the maximum likelihood occurs for $L=2.1\,L_0$ ($29$\,Gpc). Several models of tori with different side lengths have been considered in~\citep{Kunz:2005wh} with the conclusions $L>19.3$\,Gpc for $\mathbb{T}^3$ and $L>14.4$\,Gpc for $S^1\times\mathbb{R}^2$. The main result of \citep{Aurich:2007yx} is that the $3$-torus with volume $\approx 5\times 10^3\,\text{Gpc}^3$ (which corresponds to side length of $17$\,Gpc) is well-compatible with the WMAP three-year data. The WMAP seven-year data has been analyzed for detecting signatures of the so-called half-turn space \citep{Aurich:2010wf} (the only difference of the half-turn space from the $3$-torus is that one of the edges is turned by $180^{\circ}$ before identifying with the opposite edge) where the case of the $3$-torus is also considered. Out of these works, only in~\citep{Phillips:2004nc} and~\citep{Kunz:2005wh} has the full covariance matrix been analyzed. For some earlier results on these topologies see also~\citep{Starobinsky:1993yx,Sokolov:1993mk,Stevens:1993zz} and references therein.

Other methods for experimental detection of non-trivial topologies have been proposed and used to analyze the experimental data. The so-called spatial cross-correlation function method has been used in~\citep{Aurich:2008km} to analyze the five-year WMAP data for signatures of a $3$-torus topology with a result $L=3.85\, H_0^{-1}$. They have also specified the orientation of the torus. In \citep{Bielewicz:2008ga} multipole vectors have been used to analyze the five-year WMAP data with the conclusion that a torus topology is slightly preferred. The authors of~\citep{Menzies:2005cg} have looked for evidence for a $3$-torus topology by the alignment of distant objects. They have put a lower bound on the side length $L>0.9 \, L_0$. For some other approaches see also~\citep{HipolitoRicaldi:2005eh,Marecki:2004yf,Gomero:2003nm,Hajian:2003ic,Opher:2004vx,Dineen:2004ar,Gomero:2001ht,Reboucas:2000du,Gomero:1999rz,Gomero:2002ki,Gomero:1998dz}.

In this work we analyze the most recent seven-year WMAP data for signatures of the three flat topologies of space mentioned above using the full covariance matrix of temperature-temperature fluctuations. By using the symmetry groups of the spaces we construct efficient algorithms for the theoretical computation of the covariance matrix and  the likelihood function using that matrix. These algorithms can be used again as soon as the high precision data from the Planck satellite \citep{:2006uk} are released. The computation of the covariance matrix is done using a modified version of the CAMB program~\citep{Lewis:1999bs}, as discussed in Sec.~\ref{covariance_matrix}, and that of $\chi^2$  and the likelihood using the available WMAP code~\citep{Jarosik:2010iu,Larson:2010gs,Komatsu:2010fb}. We have used only the $TT$ correlations in our analysis. Including $TE$, $EE$, and $BB$ correlations is straightforward, but would quadruple the computer time needed, without much improvement in the results since these other correlations have much larger errors.

There have been speculations in the literature about the detection of a special direction in the CMB map in which the first few multipoles of temperature-temperature correlations seem to be aligned~\citep{Land:2005ad,Land:2006bn,Rakic:2007ve}. This is referred to as the ``axis of evil'' and is given by $b=60^{\circ}$, $l=-100^{\circ}$ in galactic coordinates. The topologies that we consider are not rotationally invariant, in particular $\mathbb{T}^2\times\mathbb{R}^1$, and $S^1\times\mathbb{R}^2$ have one special direction (the infinite one in $\mathbb{T}^2\times\mathbb{R}^1$, and the finite one in $S^1\times\mathbb{R}^2$), so we analyze the case where this special direction coincides with the axis of evil, to see if the axis of evil can be explained by one of these topologies. The authors of~\citep{Cresswell:2005sh} have analyzed the topology $S^1\times\mathbb{R}^2$ with the conclusion that it is not the explanation for the multipole alignment.

The work is organized as follows. In section \ref{quantum_creation} we present a slight generalization of Linde's argument for the quantum creation of compact universes. We describe the calculation of the covariance matrix and the likelihood in sections \ref{covariance_matrix} and \ref{likelihood} respectively. We present our numerical results in section \ref{summary}, and discuss the goodness of fit, and maximum likelihood confidence intervals. We have done several checks of our analysis, which are given in Sec.~\ref{sec:checks}. The possibility that our results are generated by a random fluctuation are analyzed in section \ref{sec:mc}, where we discuss Monte-Carlo skies.
The possibility of spurious effects due to a small residual CMB dipole in the data is investigated in Sec.~\ref{sec:dipole}.  We summarize in section \ref{conclusions}. Unless otherwise stated, everywhere in this work the side length of the torus is given in units of the distance to the last scattering surface $L_0$.


\section{Quantum creation of compact universes}\label{quantum_creation}

Consider the standard Einstein-Hilbert action of gravity minimally coupled to matter. Here we will be only interested in ``quantizing'' gravity, so for the matter portion we will just consider energy density $V$ without worrying about where it comes from. Then the action takes the form ($\hbar=c=1$, $M_{pl}\equiv(8\pi G)^{-1/2}=1$)
\begin{equation}\label{action1}
S=\frac{1}{2}\int \rd^4x\sqrt{-g}\left(R-2V\right)\,.
\end{equation}
Following the standard procedure to derive the Wheeler-DeWitt equation we use the ADM form of the spacetime metric \citep{Arnowitt:1962hi}
\begin{equation}\label{metric_ADM}
\rd s^2=-N^2\rd t^2+h_{ij}(\rd x^i+N^i\rd t)(\rd x^j+N^j\rd t)\,.
\end{equation}
The action can be rewritten in the form
\begin{equation}\label{action_ADM}
S=\int \rd^4x\ \mathcal{L}\,,
\end{equation}
with
\begin{equation}\label{L_ADM}
\mathcal{L}=\frac{\sqrt{h}N}{2}\left({}^3R+\frac{1}{N^2}(E_{ij}E^{ij}-E^2)-2NV\right)
\end{equation}
where ${}^3R$ is the $3$-curvature of spatial slices,
\begin{eqnarray}\label{E_ij}
E_{ij}&=&\frac{1}{2}(\dot{h}_{ij}-\nabla_iN_j - \nabla_jN_i)\,,\nn
E&=&E_i^i\,.
\end{eqnarray}

There are numerous possibilities for the spacetime manifold and it may be described by infinitely many parameters, so to be able to proceed we consider manifolds with finite homogeneous spatial slices which can be characterized by one length scale $a(t)$. In other words, we assume that locally the manifold is characterized by a Friedmann-Robertson-Walker metric while globally it can have any finite topology that is compatible with the metric. So by a suitable choice of coordinates we get in this case
\begin{eqnarray}\label{N}
N=1,\quad N_i=0\,,\\
\label{h_ij}
h_{ij}=a^2(t)k_{ij}\,,
\end{eqnarray}
where the tensor $k_{ij}$ is constant (it only depends on the choice of the manifold but does not depend on any of the coordinates). Then
\begin{equation}
E_{ij}E^{ij}-E^2=-6\left(\frac{\dot{a}}{a}\right)^2\,.
\end{equation}

Since we assumed a homogeneous spatial submanifold characterized by single length scale $a$, by dimensional analysis the volume must be proportional to $a^3$ and the curvature to $a^{-2}$. Namely,
\begin{eqnarray}\label{alpha}
\int \rd^3x\ \sqrt{h}&=&\alpha a^3\,, \\
\label{beta}
{}^3R&=&\frac{\beta}{a^2}\,,
\end{eqnarray}
where $\alpha$ and $\beta$ are dimensionless constants that depend only on the choice of the manifold. The Lagrangian then takes the form
\begin{equation}\label{lagrangian}
L=\frac{\alpha}{2}\left(a\beta-6a\dot{a}^2-2a^3V\right)\,.
\end{equation}

Now we treat $a$ as the dynamical variable describing the geometry. The canonical momentum is then
\begin{equation}\label{p_a}
p_a=\frac{\partial L}{\partial\dot{a}}=-6\alpha a\dot{a}\,,
\end{equation}
and the Hamiltonian becomes
\begin{equation}\label{hamiltonian}
H=p_a\dot{a}-L=\frac{1}{12\alpha a}\left(-p_a^2-6\alpha^2\beta a^2+12\alpha^2a^4V\right)\,.
\end{equation}

Finally, we canonically quantize, replacing $p_a$ by the operator $-i(\rd/\rd a)$ to get for the Hamiltonian
\begin{equation}\label{hamiltonian_quantum}
H=\frac{1}{12\alpha a}\left(\frac{d^2}{da^2}-6\alpha^2\beta a^2+12\alpha^2a^4V\right)\,.
\end{equation}

\begin{figure}
\includegraphics[width=8cm,bb=60 158 468 460]{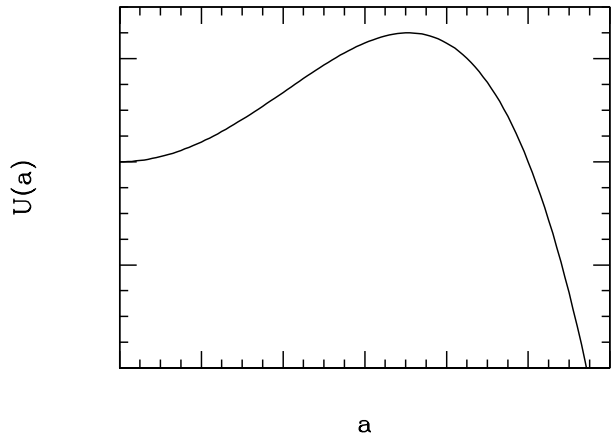}
\caption{\label{fig:U} Plot of the potential $U(a)$ for a positively curved space ($\beta>0$). The axes are in arbitrary units.}
\end{figure}

Consider now the quantum creation of the universe with zero energy. Then the wavefunction of the universe $\Psi(a)$ satisfies the analog of the Schr\"{o}dinger equation with Hamiltonian Eq.~(\ref{hamiltonian_quantum}), which is called the Wheeler-DeWitt equation. In this case it takes the form
\begin{equation}\label{wheeler_dewitt}
\left(\frac{d^2}{da^2}-6\alpha^2\beta a^2+12\alpha^2a^4V\right)\Psi(a)=0\,.
\end{equation}
The effective potential energy is
\begin{equation}\label{veff}
U(a)=6\alpha^2\beta a^2-12\alpha^2a^4V\,,
\end{equation}
and is shown in Fig.~\ref{fig:U}
which decreases to $-\infty$ for large $a$ since the second term dominates. However, for small $a$ the first term dominates, so for $\beta>0$ there is a potential barrier from $a=0$ to $\sqrt{\beta/2V}$, i.e. the universe has to first undergo tunneling before the expansion can start. This is the reason why the probability of quantum creation of positively curved spaces, which have $\beta>0$, is thought to be highly suppressed compared to flat and negatively curved spaces. The action for tunneling through the barrier is $S=a_0^3 \sqrt{V}/3$, where $a_0=\sqrt{\beta/(2V)}$ is the size of the created universe. The tunneling probability is $\propto \exp(-S)$, and is greater for smaller universes; $S\to 0,a_0 \to 0$ as $V \to \infty$. For a flat universe, $\beta=0$, and the barrier vanishes.

\section{Covariance matrix calculation}\label{covariance_matrix}

Now we turn to the calculation of correlations between CMB temperature anisotropies in the flat topologies $\mathbb{T}^3$, $\mathbb{T}^2\times\mathbb{R}^1$, and $S^1\times\mathbb{R}^2$. Locally they all look exactly like the infinite flat $\mathbb{R}^3$ so Einstein's equations and therefore the Friedmann equations are unchanged from the infinite case. The calculation for the infinite case is described in standard textbooks (for a detailed derivation see \citep{dodelson}), so let us briefly summarize that calculation and then focus on the differences between the infinite and finite universes. Essentially, one has to take the Einstein's equations that describe the interactions between gravity and all of matter and Boltzmann's equations for interactions between various types of matter (most importantly, electrons and photons) and solve for the distribution of photons today given initial conditions set by inflation. Since the temperature anisotropies in the CMB are about five orders of magnitude smaller than the background, the calculation is done using perturbation theory around the homogeneous background and keeping only first order terms. Then all of the differential equations become linear and can be treated easily in Fourier space. This is where there is a key difference: in an infinite universe the spectrum of the Fourier modes $\mathbf{k}$ is continuous, while for compactified dimensions the spectrum becomes discrete. For a torus with side lengths $L_1$, $L_2$, and $L_3$ we have $\mathbf{k}=(k_1,k_2,k_3)$,
\begin{equation}\label{k_spectrum}
k_1=\frac{2\pi}{L_1}n_1,\quad k_2=\frac{2\pi}{L_2}n_2,\quad k_3=\frac{2\pi}{L_3}n_3\,,
\end{equation}
where $n_1$, $n_2$, $n_3$ are integers (the torus is essentially a box with periodic boundary conditions). So all of the equations in Fourier space remain unchanged, all we have to worry about is integrations over $\mathbf{k}$ which have to be replaced by sums
\begin{equation}\label{replacement}
\int\frac{\rd^3k}{(2\pi)^3}\rightarrow\frac{1}{L_1L_2L_3}\sum_{\mathbf{k}}\,,
\end{equation}
over the discrete $\mathbf{k}$ values in Eq.~(\ref{k_spectrum}). The set of points Eq.~(\ref{k_spectrum}) will be referred to as the $\mathbf{k}$ grid.

The three cases studied here can be characterized by different values for $L_i$. The three-torus $\mathbb{T}^3$ has
$L_1=L_2=L_3=L$,  $\mathbb{T}^2\times\mathbb{R}^1$ has $L_1=L_2=L$, $L_3=\infty$, and finally $S^1\times\mathbb{R}^2$ has $L_1=L_2=\infty$, $L_3=L$. All three cases can be treated in a unified manner by using the integral notation, with the understanding that the integral is to be replaced by a summation if the corresponding $L_i$ is finite.

The first set of summations over $\mathbf{k}$ arises when constructing collision terms in Boltzmann's equations. However, we will not worry about these integrals for the following reason. The Boltzmann's equations are important only before the decoupling epoch, which corresponds to a redshift of about $z\sim1100$. The comoving horizon at that time was about $50$ times bigger than currently, and the current bounds on the size of the torus are of the order of the size of horizon, so at the epoch of decoupling, the size of the torus was at least about $50$ times bigger than the causally connected part. As we will see later in section \ref{summary}, the sums rapidly converge to the corresponding integrals when the topology scale is around $3$ times the radius of horizon, which implies that the effects of finiteness can be safely ignored for the epoch of decoupling (and before), and $\mathbf{k}$ can be treated as a continuous variable. All the equations are solved in Fourier space, exactly as for the infinite case. 

There is a summation over $\mathbf{k}$ when the final answer for the temperature fluctuations has to be converted from Fourier space back to real space. So let us pick up from that point in the calculation. The temperature anisotropies $\Theta(\hat{\mathbf{n}}, \mathbf{x})$ in direction $\hat{\mathbf{n}}$ at a given point $\mathbf{x}$ (chosen to be our location $\mathbf{x}_0$) are decomposed into spherical harmonics
\begin{equation}\label{ylm_decomposition}
\Theta(\hat{\mathbf{n}}, \mathbf{x}) = \sum_{l=1}^{\infty}\sum_{m=-l}^{l}a_{lm}(\mathbf{x})Y_{lm}(\hat{\mathbf{n}})\,,
\end{equation}
where the position space $a_{lm}$ coefficients are given in terms of the Fourier space temperature fluctuations  $\Theta(\hat{\mathbf{n}}, \mathbf{k})$ by
\begin{equation}\label{a_lm}
a_{lm}(\mathbf{x})=\int\frac{\rd^3k}{(2\pi)^3}e^{i\mathbf{k}\cdot\mathbf{x}}\int \rd\Omega\ Y_{lm}^*(\hat{\mathbf{n}})\Theta(\hat{\mathbf{n}}, \mathbf{k})\,.
\end{equation}
The observed CMB fluctuations are given by the correlations between the different $a_{lm}$'s,
\begin{equation}\label{M}
M_{lml'm'}\equiv \vev{ a_{lm}(\mathbf{x}_0) a_{l'm'}^*(\mathbf{x}_0) } \,.
\end{equation}

The correlations between temperature anisotropies in $k$-space are related to the initial matter power spectrum
\begin{eqnarray}\label{theta-theta}
&&\vev{ \Theta(\mathbf{k},\hat{\mathbf{n}})\Theta^*(\mathbf{k}^\prime,\hat{\mathbf{n}}^\prime) } \nn
&=&(2\pi)^3\delta^3(\mathbf{k}-\mathbf{k}^\prime)P(k)\frac{\Theta(k, \mathbf{k}\cdot\hat{\mathbf{n}})}{\delta(k)}\frac{\Theta^*(k, \mathbf{k}\cdot\hat{\mathbf{n}}^\prime)}{\delta^*(k)}\,,
\end{eqnarray}
where the matter power spectrum is defined by \citep{dodelson}
\begin{equation}\label{matter_P}
\vev {\delta(\mathbf{k})\delta^*(\mathbf{k}^\prime ) } \equiv(2\pi)^3\delta^3(\mathbf{k}-\mathbf{k}^\prime)P(k)\,.
\end{equation}

The ratios $\Theta/\delta$ on the right hand side of Eq.~(\ref{theta-theta}) do not depend on the initial conditions since the equations are linear. All of the information about initial conditions is now absorbed into $P(k)$. From Eq.~(\ref{a_lm}), (\ref{M}), and (\ref{theta-theta}) we get
\begin{widetext}
\begin{equation}\label{M1}
M_{lml'm'}=\int\frac{\rd^3k}{(2\pi)^3}P(k)\int \rd\Omega\ Y_{lm}^*(\hat{\mathbf{n}})\frac{\Theta(k, \mathbf{k}\cdot\hat{\mathbf{n}})}{\delta(k)}\int \rd\Omega^\prime\ Y_{l^\prime m^\prime }(\hat{\mathbf{n}}^\prime)\frac{\Theta^*(k, \mathbf{k}\cdot\hat{n}^\prime)}{\delta^*(k)}\,.
\end{equation}
\end{widetext}
Expanding $\Theta(k, \mathbf{k}\cdot\hat{\mathbf{n}})$ into Legendre polynomials
\begin{equation}\label{theta_expansion}
\Theta(k, \mathbf{k}\cdot\hat{\mathbf{n}})=\sum_{l}(-i)^l(2l+1)P_l(\hat{k}\cdot\hat{\mathbf{n}})\Theta_l(k)\,,
\end{equation}
and using the identity
\begin{equation}\label{legendre_identity}
\int \rd\Omega\ P_{l^\prime}(\hat{k}\cdot\hat{\mathbf{n}})Y_{lm}(\hat{\mathbf{n}})=\frac{4\pi }{2l+1}\delta_{ll^\prime} Y_{lm}(\hat{k})\,,
\end{equation}
we finally get
\begin{eqnarray}\label{M2}
&& M_{lml^\prime m^\prime }= (4\pi)^2(-i)^li^{l^\prime } \times \nn
&&\int\frac{\rd^3k}{(2\pi)^3}P(k)\frac{\Theta_l(k)}{\delta(k)}\frac{\Theta_{l^\prime }^*(k)}{\delta^*(k)} Y_{lm}^*(\hat{\mathbf{k}})Y_{l^\prime m^\prime }(\hat{\mathbf{k}})\,.
\end{eqnarray}

We first review the standard result for an infinite universe. In this case, the angular part of the integral over $\mathbf{k}$ in Eq.~(\ref{M2})
can be done analytically giving
\begin{equation}\label{M_inf}
M_{lml^\prime m^\prime }=\delta_{ll^\prime }\delta_{mm^\prime }C_l\,,
\end{equation}
with
\begin{equation}\label{C_l}
C_l=\frac{2}{\pi}\int \rd k\ k^2\, P(k)\left|\frac{\Theta_l(k)}{\delta(k)}\right|^2\,.
\end{equation}

The derivation remains the same in the finite case except that the $\mathbf{k}$ integral must be replaced by the sum Eq.~(\ref{replacement}), so instead of Eq.~(\ref{M2}) we get
\begin{eqnarray}\label{M_torus}
&& M_{lml'm'} = (4\pi)^2(-i)^li^{l^\prime} \times \nn
&& \frac{1}{L_1L_2L_3}\sum_{\mathbf{k}}P(k)\frac{\Theta_l(k)}{\delta(k)}\frac{\Theta_{l^\prime}^*(k)}{\delta^*(k)}
 Y_{lm}^*(\hat{\mathbf{k}})Y_{l^\prime m^\prime}(\hat{\mathbf{k}})\,.
\end{eqnarray}
Now we have to compute a three-dimensional sum Eq.~(\ref{M_torus}) instead of a one-dimensional integral Eq.~(\ref{C_l}) which requires much more computational time. Also, we have to calculate all matrix elements with different $l$, $m$, $l'$, $m'$ whereas in the infinite case all $l \not = l^\prime$ or $m \not = m^\prime$ (non-diagonal) elements vanish, while the diagonal ones do not depend on $m$. The reason for this is clear. In the infinite case, the problem has full rotational invariance, so that angular momentum is conserved. In the cases we consider, rotational invariance is broken. Even though rotational invariance is broken, there is still a large residual discrete symmetry group which can be used to simplify the problem, and reduce the computational time. We will refer to this residual symmetry group as $G$. For the $\mathbb{T}^2 \times \mathbb{R}^1$ and $S^1 \times \mathbb{R}^2$ cases, $G$ is the  symmetry group of a rectangular parallelepiped with two sides equal, the tetragonal group $D_{4h}$ with 16 elements, whereas for $\mathbb{T}^3$, $G$ is the symmetry group of the cube, the octahedral group $O_h$ with 24 elements.

The angular part in the sum in Eq.~(\ref{M_torus}) can be separated (this has been suggested earlier in \citep{Phillips:2004nc})
\begin{eqnarray}\label{M_torus_angular}
M_{lml^\prime m^\prime }&=&\frac{(4\pi)^2(-i)^li^{l^\prime }}{L_1L_2L_3}\sum_{k}P(k)\frac{\Theta_l(k)}{\delta(k)}\frac{\Theta_{l^\prime }^*(k)}{\delta^*(k)}\nn
&&\qquad \times \sum_{|\mathbf{k}|=k}Y_{lm}^*(\hat{\mathbf{k}})Y_{l^\prime m'}(\hat{\mathbf{k}})\,,
\end{eqnarray}
where the first sum is over all the allowed spheres in the $\mathbf{k}$ grid while the angular sum is over a fixed sphere and depends only on the choice of that sphere. 

We can simplify the computation using the discrete symmetry group $G$ of the manifolds $\mathcal{M}_{0,1,2}$.
Consider a fixed sphere of radius $k$. If the point $(\theta, \phi)$ of the sphere is on the grid, then so are $(\theta, \phi +\pi/2)$, $(\theta, \phi +\pi)$, and $(\theta, \phi + 3\pi/2)$. The angular sum over these four points is proportional to
\[
e^{i(m^\prime -m)\phi}\left(1+e^{i(m^\prime -m)\frac{\pi}{2}}+e^{i(m^\prime -m)\pi}+e^{i(m^\prime -m)\frac{3\pi}{2}}\right)\,,
\]
which is $0$ unless $m^\prime -m$ is divisible by $4$, in which case it becomes $4e^{i(m^\prime -m)\phi}$. Consider the points $(\theta, \phi)$ and $(-\theta, \phi)$, which both lie on the sphere. Since
\[
Y_{lm}(-\theta, \phi)=(-1)^{l-m}Y_{lm}(\theta, \phi)\,,
\]
the sum over those two points is $0$ unless $l+l^\prime -m-m^\prime $ is even, but $m+m^\prime $ is even if $m^\prime -m$ is divisible by $4$, so the extra condition we get is that $l^\prime -l$ has to be even (this also follows from parity). The eight points $(\pm \theta, \phi + n \pi/2)$, $n=0,1,2,3$ lie in the eight different octants, so the point $(\theta,\phi)$ can be chosen to lie in the first octant. To summarize, the angular sum is nonzero only if $l^\prime -l$ is even and $m^\prime -m$ is divisible by $4$, in which case it is equal to $8$ times the sum over one octant. Extra care is needed for points on the boundary of the octant to avoid double counting.

Consider the points $(\theta, \phi)$ and $(\theta, \pi/2-\phi)$ corresponding to swapping $n_1$ with $n_2$. Taking into account that $m^\prime -m$ is divisible by $4$, we get
\[
e^{i(m^\prime -m)\phi}+e^{i(m^\prime -m)(\frac{\pi}{2}-\phi)}=2\cos\left(\left(m^\prime -m\right)\phi\right)\,,
\]
which implies that the angular sums are real. Furthermore, $(-i)^li^{l^\prime }$ is also real for even $l^\prime -l$ and $P(k)$ and $\Theta_l(k)/\delta(k)$ are real, so the covariance matrix elements $M_{lml^\prime m^\prime }$ are all real implying $M_{lml^\prime m^\prime }=M_{l^\prime m^\prime lm}$. Also, since $Y_{l,-m}(\theta,\phi)=(-1)^mY_{lm}^*(\theta,\phi)$ and $m^\prime -m$ is divisible by $4$, we get $M_{lml^\prime m^\prime }=M_{l,-m,l^\prime ,-m^\prime }$.

$\mathbb{T}^3$ has more symmetries which can be used to further speed up the calculation for this case. For example, the sums in Eq.~(\ref{M_torus_angular}) over the spherical harmonics are the same for all $L$. Changing $L$ is a rescalling of the allowed momenta by $1/L$. Thus the angular sum for $\abs{\mathbf{k}}=k$ for a $\mathbb{T}^3$ of size $L$ is the same as the angular sum for $\abs{\mathbf{k}}=\lambda k$ for $\mathbb{T}^3$ of size $L/\lambda$. Thus the angular sums can be computed once, and then used for all values of $L$.

\begin{figure}
\includegraphics[width=8cm,bb=60 158 460 470]{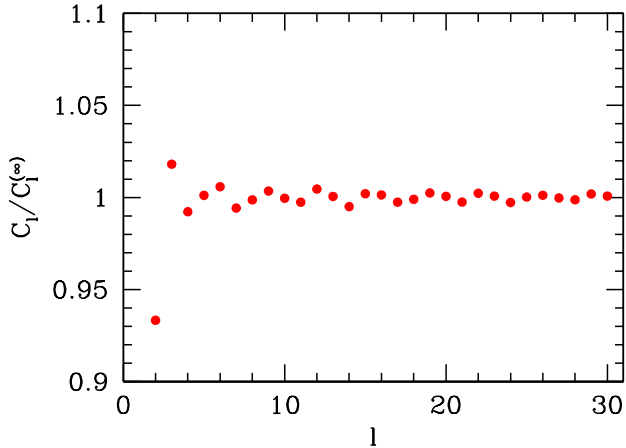}
\caption{\label{fig:cl} Plot of the ratio of $C_l$ for $\mathcal{M}_1=\mathbb{T}^2 \times \mathbb{R}^1$ with $L/L_0=1.9$, to that for infinite space $\mathbb{R}^3$.}
\end{figure}

Since the calculation of $\Theta_l(k)/\delta(k)$ is identical to the case of infinite flat universe, we use the well-known CAMB software \citep{Lewis:1999bs} (based on CMBFAST \citep{Seljak:1996is}) for that part of the calculation. Our code is essentially a modification of CAMB. It takes the sides of the torus as extra input parameters and outputs not only $C_l$ but also the complete matrix $M_{lml^\prime m^\prime }$.
For large values of $l$, the discrete sums over $\mathbf{k}$ approach the continuum result, so we only use Eq.~(\ref{M_torus_angular}) for $l \le 30$, and use the continuum result for $l > 30$. The difference between the discrete and continuum values for $M_{lml^\prime m^\prime }$ is less than 0.5\% for $l=30$. As an example, in Fig.~\ref{fig:cl}, we have plotted
the ratio of the power spectrum $C_l$ for $\mathcal{M}_1=\mathbb{T}^2 \times \mathbb{R}^1$ with $L/L_0=1.9$ to that for infinite space $\mathbb{R}^3$, where $C_l$ has been defined as 
\begin{eqnarray}
C_l &=& \frac{1}{2l +1} \sum_{m=-l}^l M_{lmlm}
\end{eqnarray}
for $\mathcal{M}_1=\mathbb{T}^2 \times \mathbb{R}^1$. The $l=2$ power is reduced by 7\%, and the ratio of power spectra oscillates
and rapidly approaches unity. The two differ by $0.1$\% at $l=30$.

\section{Likelihood calculation}\label{likelihood}

The matrix $M_{lml^\prime m^\prime }$ computed as discussed above is compared to the experimental data from the 7-year WMAP survey. Since rotational invariance is broken, we need to vary the orientation of the torus relative to axes fixed in space, to find the best fit. We do this by rotating the data relative to the torus in computing the likelihood function. We specify the orientation by three Euler angles $(\phi, \theta, \psi)$ in the following way. The axes $x,y,z$ are fixed in the coordinate frame of the CMB data, i.e.\ the observed universe, and the $x^\prime, y^\prime,z^\prime$ axes are fixed in the torus. Start with the CMB-fixed and torus-fixed axes aligned. Rotate the torus counterclockwise around the $z$-axis by angle $\phi$, then around the \emph{new} $x$-axis by angle $\theta$, then around the \emph{new} $z$-axis by angle $\psi$ to get the final torus orientation. The angles $\phi$ and $\theta$ give the orientation of the $z$-axis of the torus while the angle $\psi$ gives the orientation of the torus around its $z$-axis. We can make use of the symmetries of our topologies to speed up the calculation since various Euler angles can give equivalent orientations of the torus. Two sets of Euler angles $(\phi, \theta, \psi)$ and $(\phi^\prime , \theta^\prime , \psi^\prime )$ are equivalent if $\exists g\in G$ such that
\begin{equation}\label{equivalence_relation}
R(\phi,\theta,\psi)=R(g)\, R(\phi^\prime ,\theta^\prime ,\psi^\prime )
\end{equation}
where $R(\phi,\theta,\psi)$ is the coordinate transformation rotation matrix corresponding to $(\phi, \theta, \psi)$ and $R(g)$ is that corresponding to the discrete element $g$. This defines an equivalence relation on the set of all possible Euler angles. We take a uniform grid on all possible angles, then divide that grid into equivalence classes according to Eq.~(\ref{equivalence_relation}) and take one representative from each class. We have scanned over $\sim 4000$ inequivalent angles.

Different orientations of the torus were considered in the previous analysis of first-year WMAP data for $\mathbb{T}^3$~\citep{Phillips:2004nc}, but they only considered a uniform grid on the range $0\leq\phi,\theta,\psi\leq\pi/2$. Note that this does not cover all possible orientations of $\mathbb{T}^3$. The first two angles describe the orientation of the $z$-axis and by their assumption on the range of $\phi,\theta,\psi$, the $z$-axis always lies in the first octant. However taking into account all the symmetries of the cube there are $6$ equivalent axes that can play the role of the $z$-axis, the $\pm x$, $\pm y$, and $\pm z$ axes, while there are $8$ octants. In other words, there are possible orientations of the cube for which none of the $6$ axes lies in the first octant.

After choosing a torus orientation, we calculate the likelihood in the real space of orientations on the last scattering surface. For $N_p$ pixels the likelihood function is given by\footnote{To distinguish from length scale $L$, the likelihood is denoted by $\mathcal{L}$ everywhere in this work.} \citep{dodelson}
\begin{equation}\label{likelihood_def}
\mathcal{L}=\frac{1}{(2\pi)^{N_p/2}(\det C)^{1/2}}\exp\left(-\frac{1}{2}\Delta^T C^{-1}\Delta\right)\,,
\end{equation}
and the $\chi^2$ function by
\begin{equation}\label{chisq_def}
\chi^2=\Delta^T C^{-1}\Delta \,,
\end{equation}
where $\Delta_i$ is the vector of pixels and $C_{ij}$ is the covariance matrix that also includes the noise. The indices $i,j$ label the different pixels, which are in directions $\hat\mathbf{n}_{i,j}$ on the sky. 

We now describe how to convert the matrix $M_{lml^\prime m^\prime }$ to the covariance matrix $C_{ij}$. The temperature fluctuation measured in a pixel $i$ is given by~\citep{dodelson}
\begin{equation}\label{beam_pattern}
\Theta_i=\int \rd\hat{\mathbf{n}}\ \Theta(\hat{\mathbf{n}})B_i(\hat{\mathbf{n}})\,,
\end{equation}
where $B_i$ is the beam pattern at the pixel $i$ and is specific to the experiment. Usually the beam patterns have the same shape for every pixel and are axially symmetric around the center of the pixel, as is the case for WMAP, so if we denote the direction to the center of the pixel by $\hat{\mathbf{n}}_i$ then the beam pattern can be decomposed into spherical harmonics
\begin{equation}\label{B_decomposition}
B_i(\hat{\mathbf{n}})=\sum_{lm}B_lY_{lm}(\hat{\mathbf{n}}_i)Y_{lm}^*(\hat{\mathbf{n}})\,.
\end{equation}
Using Eq.~(\ref{ylm_decomposition}) to decompose $\Theta(\hat{\mathbf{n}})$ into spherical harmonics, we get for the theoretical covariance matrix
\begin{equation}\label{C_ij}
C_{ij}\equiv \vev{\Theta_i\Theta_j}=\sum_{lml^\prime m^\prime }M_{lml^\prime m^\prime }B_lB_{l^\prime }Y_{lm}(\hat{\mathbf{n}}_i)Y_{l^\prime m^\prime }^*(\hat{\mathbf{n}}_j)\,.
\end{equation}
In computing $C_{ij}$, we have to vary the orientation of the torus relative to the sky. In implementing the Euler angle rotation, one can compute the $M_{l m l^\prime m^\prime}$ matrix in the torus-fixed coordinate system, so that it remains unchanged as the Euler angles are varied. The pixel directions $\mathbf{n}_i$ are changed to
$\mathbf{n}_i \to R(\phi,\theta,\psi) \,\mathbf{n}_i$. Equivalently, one can work in the CMB-fixed coordinate system, and rotate the torus, which gives $M_{l m l^\prime m^\prime}$ transformed by the angular momentum rotation matrices,
\begin{eqnarray}
M_{l m l^\prime m^\prime} &\to & \sum_{n,n^\prime} M_{l n l^\prime n^\prime} D^{(l)*}_{n m}(R) D^{(l)}_{n^\prime m^\prime}(R)\,.
\end{eqnarray}
Note that in the infinite universe case, Eq.~(\ref{M_inf}) holds, and the result Eq.~(\ref{C_ij}) simplifies to
\begin{equation}\label{C_ij_inf}
C_{ij}=\sum_l\frac{4\pi}{2l+1}B_l^2\, C_l\,P_l(\hat{\mathbf{n}}_i\cdot\hat{\mathbf{n}}_j)\,,
\end{equation}
independent of the rotation $R(\phi,\theta,\psi)$.

The computation of the covariance matrix using Eq.~(\ref{C_ij}) is more involved than the infinite case, Eq.~(\ref{C_ij_inf}), so the likelihood  calculations require far more computer time than the conventional case. $C_{ij}$ must be recalculated for each set of Euler angles. There are 458403 independent elements in $M_{l m l^\prime m^\prime}$ for $2 \le l \le 30$ of which 57840 satisfy the $l\equiv l^\prime\, (\text{mod}\, 2), m\equiv m^\prime\, (\text{mod} \, 4)$ condition, and 2482 values for each of the indices $i$ and $j$.  The slowest step in the computation is evaluating the sums on $l,m,l^\prime,m^\prime$ in Eq.~(\ref{C_ij}) for all values of $\left\{ i,j\right\}$.

An Euler angle rotation of the sky maps points on the sphere to rotated points on the sphere. For infinitesimal pixels, this corresponds to a reshuffling of the pixels, i.e.\ if pixel $i$ at $\mathbf{n}_i$ is mapped by the rotation to $\mathbf{n}_j$, then pixel $i$ $\to$ pixel $j$. An exact reshuffling of pixels would greatly simplify the computation --- instead of recomputing $C_{ij}$, one could simply permute the indices on $C_{ij}$ to get the transformed matrix. In particular, $\det C$ would remain invariant under this transformation.

The WMAP pixels have been chosen using the HEALPix grid~\citep{Gorski:2004by}. The pixels are chosen to lie along lines of constant lattitude, and they have equal solid angles. This implies that the spacing of the pixels varies as a function of lattitude. As a result one cannot treat rotations of the sky as a pixel reshuffling transformation. One can approximate the rotations by a pixel transformation by mapping the rotated pixel to the one closest to it in the HEALPix grid. The likelihood computed using this method differs from the exact result using Eq.~(\ref{C_ij}), and is not accurate enough for our purposes. The above approximate relation between rotations and pixel permutations does, however, explain why $\det C$ is approximately independent of the Euler angles.

For a finite universe, one has to use Eq.~(\ref{C_ij}) with the value for $M_{lml^\prime m^\prime }$ computed as described in Sec.~\ref{covariance_matrix}. The finiteness of the universe only affects the large-scale anisotropies, so the difference between the infinite and finite cases goes to zero with increasing $l$. For that reason we will look only at low-$l$ portion of anisotropies, $l \le 30$, and use the infinite manifold result Eq.~(\ref{C_ij_inf}) for $l >30$. We calculate $\chi^2$ and the likelihood $\mathcal{L}$ using a modification of the likelihood code provided by the WMAP team~\citep{Jarosik:2010iu,Larson:2010gs,Komatsu:2010fb} as a function of the new parameters ($L$, $\phi$, $\theta$, $\psi$). Since we are interested only in low-$l$ effects we use the low-resolution portion of the likelihood code. We use the experimental data in the exact same form as provided by the WMAP team without any further modifications. The temperature map used is the smoothed and degraded ILC map with the Kp2 mask applied to remove the galactic plane and strong point sources. The map originally has 3072 pixels, but only 2482 are left after the mask.

Ideally, one would have to do a fit to the experimental data varying the four new parameters ($L$, $\phi$, $\theta$, $\psi$) in addition to all the other cosmological parameters. The cosmological parameters affect the whole spectrum of anisotropies while only the low-$l$ part of the spectrum is affected by the new parameters, so we fix the other cosmological parameters at their best-fit values as given by the  seven-year WMAP data~\citep{Komatsu:2010fb} and only vary the new parameters. The values of the cosmological parameters that we use are~\citep{Komatsu:2010fb} $100\Omega_bh^2=2.227$, $\Omega_ch^2=0.1116$, $\Omega_{\Lambda}=0.729$, $n_s=0.966$, $\tau=0.085$, $\Delta_R^2(0.002\,\text{Mpc}^{-1})=2.42\times10^{-9}$.

\newpage

\section{Results}\label{summary}

We have computed the likelihood and $\chi^2$ for the three cases, $\mathcal{M}_0=\mathbb{T}^3$,
$\mathcal{M}_1=\mathbb{T}^2 \times \mathbb{R}^1$ and $\mathcal{M}_2=S^1 \times \mathbb{R}^2$, for different values of $L/L_0$ as a function of the Euler angles. $L/L_0$ ranges between a minimum value of $0.5-1.0$  depending on the manifold, and a maximum value of $L/L_0=2.6$,  in steps of $L/L_0=0.1$. By the time $L/L_0=2.6$, the results are almost identical to the flat-space case $\mathcal{M}_{\infty}=\mathbb{R}^3$.  For the statistical analysis discussed later in this section, we have used interpolation to construct a smooth function of $L$.

The relation between likelihood and $\chi^2$ is
\begin{eqnarray}\label{reln}
-2 \ln \mathcal{L} &=& \chi^2 + \ln \det C/C_f + \ln \det (2 \pi C_f)\,
\end{eqnarray}
where $C_f$ is a fiducial covariance matrix used by the WMAP collaboration. $C_f$ is independent of $L$ and the Euler angles, and drops out of all likelihood ratios. $\chi^2$ and $-2\ln\mathcal{L}$  differ by $\ln \det C/C_f$ (up to an irrelevant constant). The likelihood for $\mathcal{M}_\infty$, three-dimensional flat space, will be denoted by $\mathcal{L}_\infty$, and is $\mathcal{L}_\infty=3573.4$

\begin{figure}
\includegraphics[width=8cm,bb=50 158 460 470]{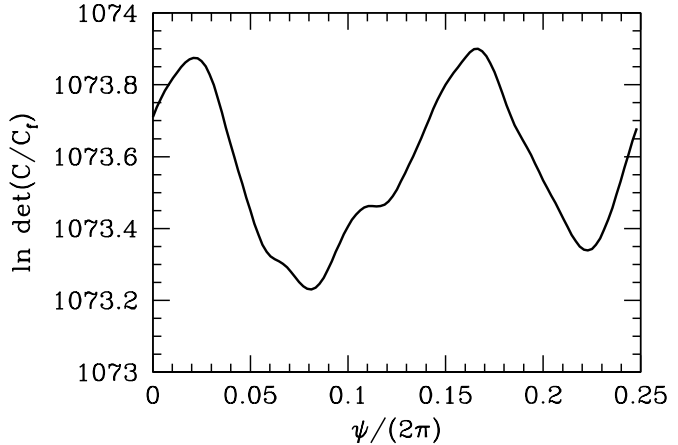}
\caption{\label{det} Plot of $\ln \det C/C_f$ against the Euler angle $\psi$  for fixed $\phi,\theta$ for the $\mathcal{M}_0$ topology at $L/L_0=1.8$,
}
\end{figure}

As noted earlier, for fixed $L/L_0$, $\ln \det C/C_f$ varies weakly with the Euler angles. In Fig.~\ref{det}, we have plotted the variation of $\ln \det C/C_f$ as a function of $\psi$, for fixed values of the $\phi,\theta$, at $L/L_0=1.8$. The overall variation of $\ln \det C/C_f$ against $\psi$ is less than unity.

\begin{figure}
\includegraphics[width=8cm,bb=60 163 460 470]{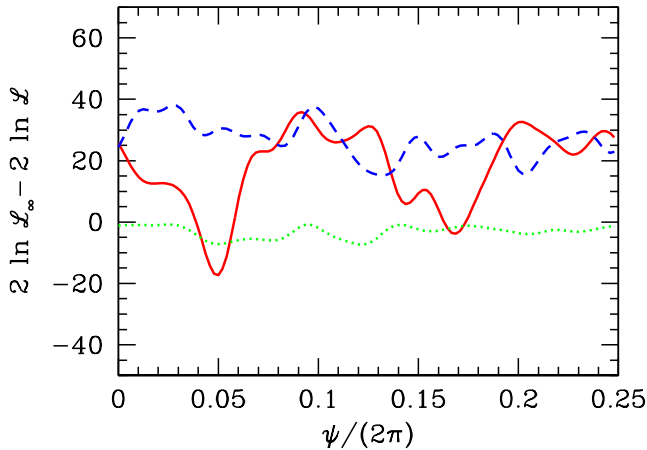}
\caption{\label{likvar} Plot of $2 \ln \mathcal{L}_\infty -2 \ln \mathcal{L}$ against the Euler angle $\psi$ for fixed $\phi,\theta$ for the $\mathcal{M}_0$ topology. The solid red and dashed blue curves are for two different values of $\phi,\theta$ at $L/L_0=1.8$, and the dotted green curve is for $L/L_0=2.2$. The solid red curve is for $L/L_0=1.8$ with $\phi,\theta$ fixed to be the best fit values, the dashed blue  curve is for $L/L_0=1.8$ with $\phi,\theta$ fixed in a random direction, and the dotted green curve is for $L/L_0=2.2$ with $\phi,\theta$ fixed to be the best fit values.
}
\end{figure}

$-2 \ln \mathcal{L}$ (and hence $\chi^2$) has a strong variation with Euler angles at fixed $L/L_0$. In Fig.~\ref{likvar}, we have shown plots of the variation of $2 \ln \mathcal{L}_\infty-2 \ln \mathcal{L}$ with Euler angle $\psi$ for fixed $\phi,\theta$ for the $\mathcal{M}_0$ topology. $\mathcal{L}_\infty$ is independent of the Euler angles. The solid red curve has been chosen to have $L/L_0=1.8$, and $\phi,\theta$ values that maximize the likelihood at this value of $L/L_0$. There is a large variation of $-2 \ln \mathcal{L}$ with the remaining Euler angle $\psi$, and the global minum of $-2 \ln \mathcal{L}$ is $2 \ln \mathcal{L}_\infty-2 \ln \mathcal{L}=-17.2$ at $\psi /(2\pi) \approx 0.05$. The strong dependence of $-2 \ln \mathcal{L}$ on orientation makes it difficult to find the true global minimum of the $-2 \ln \mathcal{L}$  and  $\chi^2$ functions. We have done a scan over all Euler angles with a spacing of $0.05 \pi$, to identify valleys, followed by a finer scan to find the mininum. By comparing our numerical minimum with the next best point, we can estimate the uncertainty in our minimum $-2 \ln \mathcal{L}$ and  $\chi^2$ values at less than $0.5$. The dashed blue curve in Fig.~\ref{likvar} is also for $L/L_0=1.8$, but with $\phi,\theta$ fixed at  random values, rather than those for which $-2 \ln \mathcal{L}$  vs. $\psi$ passes through the global minimum. There is  still considerable dependence as one varies the third angle $\psi$, but the dependence is much weaker than for the solid red curve. The dependence of $-2 \ln \mathcal{L}$  drops rapidly with increasing $L/L_0$. For $L/L_0=2.2$, the dotted green curve in the figure, the overall variation is about 6.5.
\begin{figure}
\includegraphics[width=8cm,bb=60 163 450 710]{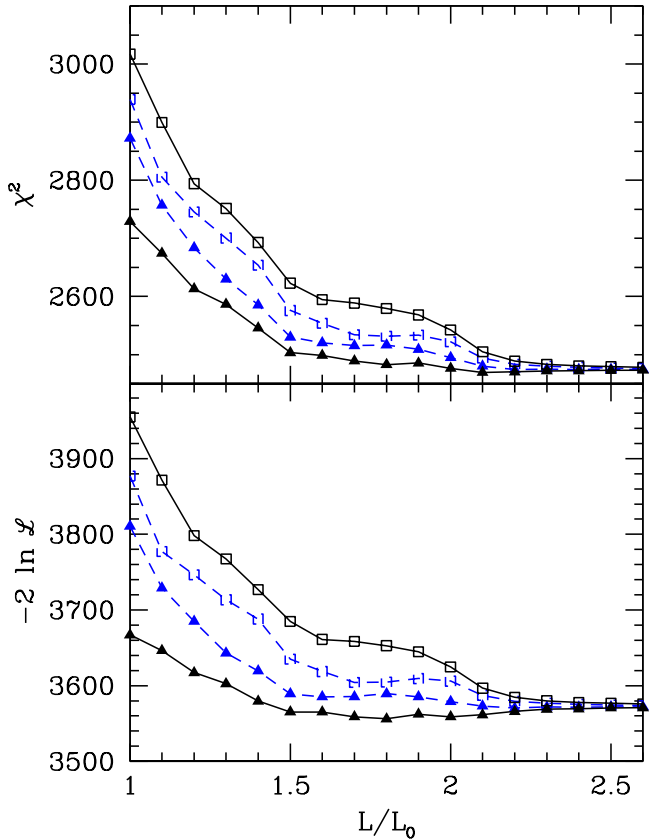}
\caption{\label{plot0} Plot of $\chi^2$ and $-2 \ln \mathcal{L}$ against $L/L_0$ for $\mathcal{M}_0=\mathbb{T}^3$ for different Euler angles $(\phi,\theta,\psi)$. The lower solid black curve (solid triangles) is the minimum $\chi^2$ (or $-2 \ln \mathcal{L}$), and the upper solid black curve (open squares) is the maximum $\chi^2$ (or $-2 \ln \mathcal{L}$). The lower dashed colored curve (solid triangles) and upper dashed colored curve (open squares) are the minimum and maximum $\chi^2$ (or $-2 \ln \mathcal{L}$) with a symmetry axis of the manifold restricted to point along the axis of evil.
}
\end{figure}
\begin{figure}
\includegraphics[width=8cm,bb=60 163 450 710]{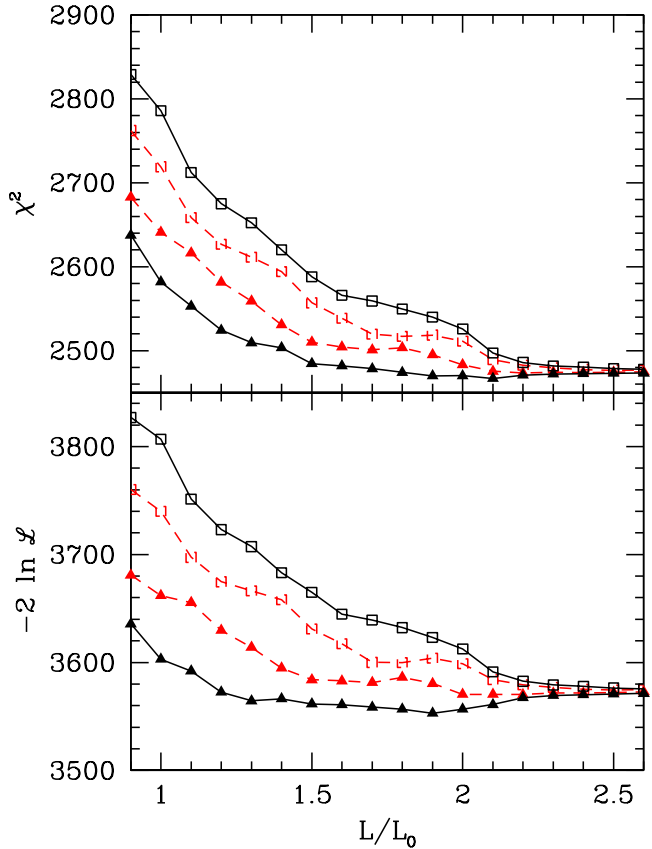}
\caption{\label{plot1} Plot of $\chi^2$  and $-2 \ln \mathcal{L}$ against $L/L_0$ for $\mathcal{M}_1=\mathbb{T}^2 \times \mathbb{R}^1$ for different Euler angles $(\phi,\theta,\psi)$. See the Fig.~\ref{plot0} caption.}
\end{figure}
\begin{figure}
\includegraphics[width=8cm,bb=60 163 450 710]{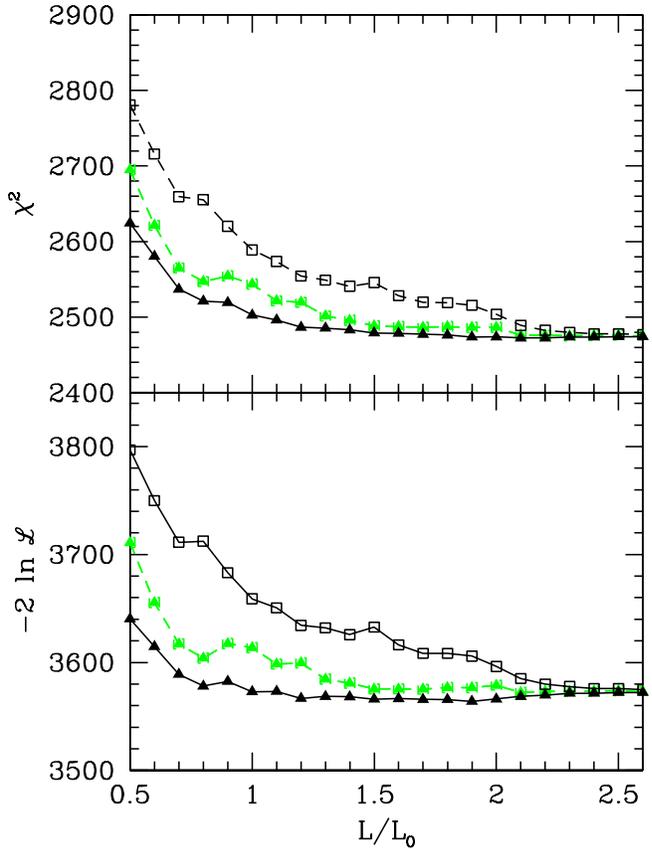}
\caption{\label{plot2} Plot of $\chi^2$  and $-2 \ln \mathcal{L}$ against $L/L_0$ for $\mathcal{M}_2=S^1 \times \mathbb{R}^2$ for different Euler angles $(\phi,\theta,\psi)$. The two colored curves are almost on top of each other.
See the Fig.~\ref{plot0} caption.}
\end{figure}

The plot of $\chi^2$ and $-2 \ln \mathcal{L}$ against $L/L_0$ is given in Figs.~\ref{plot0}, \ref{plot1}, \ref{plot2} for the three cases, $\mathcal{M}_0=\mathbb{T}^3$, $\mathcal{M}_1=\mathbb{T}^2 \times \mathbb{R}^1$ and $\mathcal{M}_2=S^1 \times \mathbb{R}^2$, respectively. We have plotted the maximum and minimum of $\chi^2$  over all possible orientations of the torus at each value of $L$. There is a significant variation in $\chi^2$ as a function of orientation, as noted earlier, and some orientations are strongly preferred over others. In each plot, $\chi^2$ ranges between the uppermost and lowermost solid black curves, as one varies the orientation of the manifold by varying the Euler angles $(\phi,\theta,\psi)$. For the smallest values of $L/L_0$, $\Delta \chi^2$ between the worst and best orientations is 377, 191, and 156 for $\mathcal{M}_{0,1,2}$, respectively. As $L/L_0$ increases, the effect of a compactified direction decreases. By the time $L/L_0=2.6$, the fit results are very close to the case of the infinite manifold $\mathbb{R}^3$, and $\Delta \chi^2 \le 4$ for the different orientations.

We have been unable to find any pattern to the best-fit orientation $\phi,\theta,\psi$ of the torus as a function of $L/L_0$. We have examined the possibility that the manifolds we consider are aligned along the axis of evil. To do this, we have chosen the preferred axis of the manifold (the $z$-axis for $\mathbb{T}^3$, the $\mathbb{R}$ direction for $\mathbb{T}^2 \times \mathbb{R}^1$ and the $S^1$ direction for $S^1 \times \mathbb{R}^2$) to point along the axis-of-evil direction $b=60^{\circ}$, $l=-100^{\circ}$ in galactic coordinates, and allowed for arbitrary rotations of the manifold around this direction. All the angles are varied with step $\pi/100=1.8^{\circ}$. This gives a subset of all the orientations we have considered, and the $\chi^2$ range has been plotted as the dashed colored curves in the figure. The colored curves lie between the black curves (as they must), but they do not lie towards the best-fit $\chi^2$ line. This shows that there is nothing in our computation that picks out the axis-of-evil as a preferred direction.

We use a goodness-of-fit test to see whether we can rule out the hypothesis that the universe has topology $\mathcal{M}_i$ of size $L$. \begin{table}
\begin{eqnarray*}
\begin{array}{cccc}
\text{C.L.} & \mathcal{M}_0 & \mathcal{M}_1 & \mathcal{M}_2 \\
\hline
68 \% & 1.47 & 1.24 & 0.92 \\
90 \% & 1.33 & 1.02 & 0.61 \\
95 \% & 1.24 & 0.96 & 0.56
\end{array}
\end{eqnarray*}
\caption{\label{tab:1} Limits on $L/L_0$ using the $\chi^2$ goodness-of-fit test. Values less than those in the table are excluded at the confidence level given in the first column.}
\end{table}
The minimum $\chi^2$ values are $\chi^2=2469,\ 2467,\ 2472$ at $L/L_0=2.1,\ 2.1,\ 2.2$ for $\mathcal{M}_{0,1,2}$, respectively. Previous studies have also found an indication of a dip in $\chi^2$ around $L/L_0 \sim 2.1$~\citep{Phillips:2004nc}. Using the computed values of $\chi^2$ we have the limits given in Table~\ref{tab:1}.

\begin{figure}[tbp]
\includegraphics[width=8cm,bb=70 153 530 540]{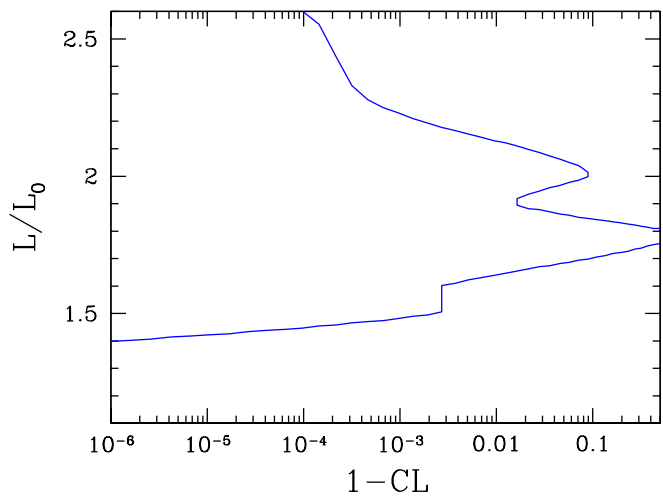}
\caption{\label{contour0} Plot of the $L/L_0$ confidence interval as a function of the confidence level for $\mathbb{T}^3$. }
\end{figure}

\begin{figure}[tbp]
\includegraphics[width=8cm,bb=70 153 530 540]{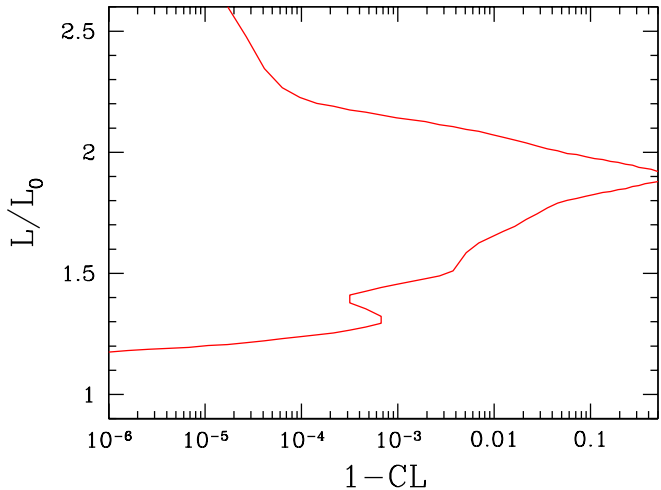}
\caption{\label{contour1} Plot of the $L/L_0$ confidence interval as a function of the confidence level for $\mathbb{T}^2 \times \mathbb{R}^1$. }
\end{figure}

\begin{figure}[tbp]
\includegraphics[width=8cm,bb=70 153 530 540]{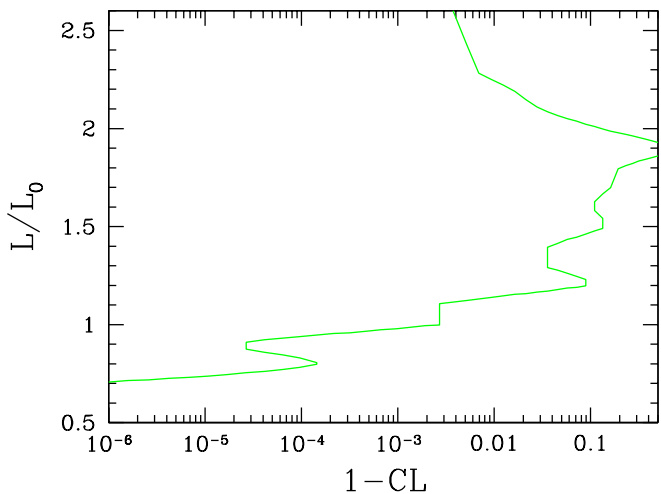}
\caption{\label{contour2} Plot of the $L/L_0$ confidence interval as a function of the confidence level for $S^1 \times \mathbb{R}^2$. }
\end{figure}

We estimate confidence intervals for $L/L_0$ using likelihood ratios~\citep{Cash:1979vz,Eadie:1971fk}\citep[\S 33]{pdg}.  Maximum likelihood confidence intervals are exact if the distribution is Gaussian. For non-Gaussian distributions, they have an error of order $1/\sqrt n$, where $n$ is the number of data points~\citep{Wilks:1938fk}. The maximum likelihood value is at $L/L_0=1.8, 1.9, 1.9$ for $\mathcal{M}_{0,1,2}$, respectively. The confidence intervals for $L/L_0$ are determined by using $\Delta \ln\mathcal{L}$, the difference of $\ln\mathcal{L}$ from the value that maximizes the likelihood function~\citep{Cash:1979vz,Eadie:1971fk}. Plots of the $L/L_0$ confidence intervals as a function of $1-\alpha$, where $\alpha$ is the confidence level, are shown in Figs.~\ref{contour0}, \ref{contour1}, \ref{contour2}. The data shows a preference for a finite universe with size $L/L_0 \sim 1.9$ corresponding to $L \sim 27$\, Gpc. The allowed $L$ range extends to $L/L_0 \ge 2.6$ at a confidence level $\alpha=10^{-4}$ for $\mathcal{M}_0$, $2 \times 10^{-5}$ for $\mathcal{M}_1$ and $4 \times 10^{-3}$ for $\mathcal{M}_2$. Thus the data show evidence for a finite universe at a confidence level $\alpha=2 \times 10^{-5}$ for the $\mathbb{T}^2 \times\mathbb{R}^1$ topology.
The 95\% confidence intervals are $L/L_0 \in \left[1.7,2.1\right], \left[1.8,2.0\right], \left[1.2,2.1\right]$ for $\mathcal{M}_{0,1,2}$, respectively.

We have scanned over $\sim 4000$ different orientations for each value of $L$, with a finer scan near the minima, so that the error in $\chi^2$  and $-2 \ln \mathcal{L}$ is $\le 0.5$.  The difference in $-2 \ln \mathcal{L}$ between its value at $L\to \infty$ and its minimum value (which occurs at $L/L_0=1.9$ for $\mathcal{M}_1$)  is $20.4$, which is well outside possible numerical errors. Note that the main numerical uncertainty is finding the true minimum of $-2 \ln \mathcal{L}$ for finite values of $L$. The minimum value of $-2 \ln \mathcal{L}$ has been determined with an accuracy $\le 0.5$. The actual difference in likelihoods between finite and infinite $L$ can only be greater than what we have found. There is an indication that a finite universe fits the data better than an infinite one. However,  the ``standard'' $5\sigma$-criterion for a discovery, corresponding to a confidence level $\alpha=5.7 \times 10^{-7}$, includes the value $L = \infty$. 

The Euler angles for the best fit case $\mathcal{M}_1$ with $L/L_0=1.9$ are $(\phi=21^\circ\pm2^\circ,\theta=53^\circ\pm2^\circ,\psi=61^\circ\pm2^\circ)$ which corresponds, for the infinite direction,  to $(b=37^\circ\pm2^\circ,l=291^\circ\pm2^\circ)$ in galactic coordinates and $(\alpha=182^\circ \pm 2^\circ, \delta=-25^\circ\pm 2^\circ)$ in J2000 equatorial coordinates.  This is  close to the direction $(b=30^\circ \pm 2^\circ, l=276^\circ \pm 3^\circ)$ of the velocity of the Local Group inferred from the CMB dipole~\citep{Lineweaver:1996qw}. We discuss the possibility that our signal is due to a dipole contamination in Sec.~\ref{sec:dipole}.

We can compare the probabilities that the universe has topology $\mathcal{M}_{0,1,2,3}$. We use the Bayesian prior that
there are four discrete choices, $\mathcal{M}_0$ with $L/L_0=1.8$, $\mathcal{M}_1$ with $L/L_0=1.9$, $\mathcal{M}_2$ with $L/L_0=1.9$, or the infinite case $\mathcal{M}_\infty$, each of which is equally probably. Since we are comparing four discrete cases, there is no ambiguity due to choice of measure in choosing an equiprobable prior. Then using the likelihood ratios gives the posterior probabilities
\begin{eqnarray*}
p\left(\mathcal{M}_0\right) &=& 0.04,\\
p\left(\mathcal{M}_1\right) &=& 0.96,\\
p\left(\mathcal{M}_2\right) &=& 2 \times 10^{-5},\\
p\left(\mathcal{M}_\infty\right) &=& 1.3 \times 10^{-9}\,.
\end{eqnarray*}
The probability of the infinite universe $\mathcal{M}_\infty$ is very small, and the most probable topology is $\mathcal{M}_1=\mathbb{T}^2 \times \mathbb{R}^1$, with two dimensions compactified, and one infinite.

\subsection{Fisher Information}

The Fisher information can  be used to compute the variance $V$ of the length $L$ determined using the maximum likelihood method. The Fisher information is given by
\begin{eqnarray}
V^{-1} &=& \frac12 \vev{\left(\frac{\partial h}{\partial L}\right)^2} = \text{Tr}\, C \frac{\partial C^{-1}}{\partial L} C
\frac{\partial  C^{-1}}{\partial L} \nn
&=& \text{Tr}\, C^{-1} \frac{\partial C}{\partial L} C^{-1}
\frac{\partial  C}{\partial L}\,.
\end{eqnarray}
Using the covariance matrix $C$ for the $\mathcal{M}_1$ toplogy with $L/L_0=1.9$ gives
\begin{eqnarray}
V^{-1} &=& 3.3 \times 10^3
\label{eq43}
\end{eqnarray}
so that the error estimate for $L/L_0$ is $\sqrt{V}=0.017$.  The Fisher information error Eq.~(\ref{eq43}) corresponds to using a quadratic approximation to the likelihood function about its minimum to determine the error, and gives a smaller error than that obtained earlier using the exact likelihood function.

\section{Checks}\label{sec:checks}

We have been unable to find a simple explanation for the better fit due to a finite topology. However, there are some possibilities which we can test.

The measured cosmic microwave background anisotropy has a smaller value for the quadrupole power $C_2$ than the theoretical expectation value. There is a large cosmic variance in $C_2$, so this is not a discrepancy between theory and experiment. Fig.~\ref{fig:cl} shows that the predicted value of $C_2$ for a finite universe is reduced from the infinite universe value. The greater likelihood for a finite universe is not due to lowering the value of $C_2$. We have checked this by determining the likelihood using $M_{lml^\prime m^\prime}$ for the finite case, but with the $l=l^\prime=2$ values replaced by their values for the infinite universe. For $\mathcal{M}_1$ with $L/L_0=1.9$, $-2 \ln \mathcal{L}$ increases by $0.43$, which is much less than the $20.4$ difference in $-2 \ln \mathcal{L}$ from the infinite universe.

\begin{figure}
\includegraphics[width=8cm,bb=60 163 460 470]{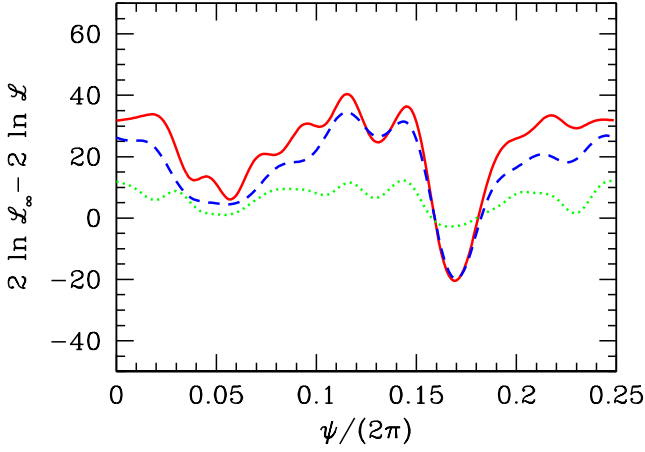}
\caption{\label{trunc1} Plot of $2\ln \mathcal{L}_\infty-2 \ln \mathcal{L}$ against the Euler angle $\psi$ for fixed $\phi,\theta$ for the $\mathcal{M}_1$ topology for $L/L_0=1.9$. The solid red curve uses the full matrix $M_{lml^\prime m^\prime}$,  the dashed blue curve uses the matrix truncated to $5 \le l,l^\prime \le 20$, and the dotted green curve uses the  matrix $M_{lml^\prime m^\prime}\delta_{ll^\prime}$, retaining only the part diagonal in $l$.
}
\end{figure}

As another test, we have computed the likelihood for the $\mathcal{M}_1$ topology for $L/L_0=1.9$ by using a truncated  $M_{lml^\prime m^\prime}$ matrix. The truncated matrix is constructed by using $M_{lml^\prime m^\prime}$ for the finite topology for $5 \le l,l^\prime \le 20$, and using $M_{lml^\prime m^\prime}$ for the infinite universe, i.e.\ $M_{lml^\prime m^\prime}=C_l \delta_{l l^\prime} \delta_{m m^\prime}$, for $l$ and or $l^\prime$ outside this range. A plot of the likelihood as a function of the Euler angle $\psi$ for this truncated $M_{lml^\prime m^\prime}$ is plotted as the dashed blue curve in Fig.~\ref{trunc1}.  This can be compared with the likelihood curve using the full $M_{lml^\prime m^\prime}$ for the finite topology, shown as the solid red curve.  The dip in the likelihood difference to $-20.4$ is the signal that the finite topology is a better fit than the infinite universe. The plot for the truncated matrix is similar to that for the full matrix, except that the small-angle fluctuations have been smoothed out, as is to be expected since higher $l$ terms have been dropped. Note that the dip in $2\ln \mathcal{L}_\infty-2 \ln \mathcal{L}$ is very similar in both cases, and the minimum of $2\ln \mathcal{L}_\infty-2 \ln \mathcal{L}$ is nearly the same. This shows that the effect we find is not due to the low-$l$ modes (quadrupole, octupole), and is also not an edge effect as a result of only using $l \le 30$ in the compuation. For $5 \le l,l^\prime \le 20$, $M_{lml^\prime m^\prime}$ has 86736 elements of which 11056 satisfy the $l\equiv l^\prime\, (\text{mod}\, 2), m\equiv m^\prime\, (\text{mod} \, 4)$ condition and are non-zero.

The off-diagonal elements in $M_{lml^\prime m^\prime}$ are important for the calculations. We have also computed the likelihood by retaining only the elements which are diagonal in $l$, i.e.\ using $M_{lml^\prime m^\prime}\delta_{l l^\prime}$. This drops the elements in $M_{lml^\prime m^\prime}\delta_{l l^\prime}$ which are off-diagonal in $l$, while retaining the elements which are off-diagonal in $m$ for a given $l$. The likelihood with this matrix is the dashed green curve in Fig.~\ref{trunc1}. With this matrix, the likelihood deviates much less from the infinite universe, and the dip near $\psi/ (2\pi) \approx 0.17$ is much less pronounced.

\section{Monte-Carlo Skies}\label{sec:mc}

The results of the previous section were obtained using a likelihood analysis of the WMAP7 data. One can study whether the better fit of a  finite topology is due to a statistical fluctuation. Since the big-bang is not a repeatable experiment, this must be done by 
generating random Monte-Carlo data for the pixels $\Delta_i$, and redoing the analysis for this Monte-Carlo data. To actually do this numerically is beyond the computing power we have available. Luckily, for the problem at hand, we can analyze the Monte-Carlo problem analytically. 

Assume that the pixels $\Delta_i$ are generated by the covariance matrix $C_\infty$ for an infinite universe, so that the probability distribution is
\begin{eqnarray}
p(\Delta) &=& \frac{1}{\sqrt{\det(2\pi C_\infty)}}\exp \left(- \frac12 \Delta^T C_\infty^{-1} \Delta\right)\,.
\label{dist}
\end{eqnarray}
The likelihood function computed using $\Delta_i$ and covariance matrix $C$ (of a finite universe) is
\begin{eqnarray}
-2 \log \mathcal{L} &=& \Delta^T C^{-1} \Delta + \ln \det (2\pi C)\,,
\label{eq10}
\end{eqnarray}
and the likelihood constructed using the covariance matrix $C_\infty$ of the infinite universe is
\begin{eqnarray}
-2 \log \mathcal{L}_\infty &=& \Delta^T C^{-1}_\infty \Delta + \ln \det (2\pi C_\infty)\,.
\label{eq10a}
\end{eqnarray}
Let
\begin{eqnarray}
h &\equiv& \left(-2 \log \mathcal{L}_\infty\right)-\left(-2 \log \mathcal{L} \right)\,,
\label{eq45}
\end{eqnarray}
be the difference of the two log-likelihoods. In our analysis, we found $h=20.4 > 0$, so that the finite universe was more likely than the infinite universe. The average value of $h$ over Monte-Carlo data can be computed using Eq.~(\ref{dist}) and Eq.~(\ref{eq45}). The two-point function is
\begin{eqnarray}
\vev{\Delta_i \Delta_j} &=& \left(C_\infty\right)_{ij}\,,
\end{eqnarray}
so that
\begin{eqnarray}
\vev{h} &=& N-\text{Tr}\, C^{-1} C_\infty + \ln \det (C_\infty )- \ln \det (C) \,.
\label{eq47}
\end{eqnarray}
It is convenient to define the symmetric matrix
\begin{eqnarray}
S=C_\infty^{1/2} C^{-1} C_\infty^{1/2}\,,
\end{eqnarray}
which is a positive matrix since $C$ and $C_\infty$ are positive matrices, and has eigenvalues $s_i >0$. In terms of $S$,
\begin{eqnarray}
\vev{h} &=& N-\text{Tr}\, S + \ln \det S \nn
&=& \sum_i \left[1-s_i + \ln s_i \right] \,.
\end{eqnarray}
The function $1 - s  + \ln s \le 0$
with its maximum at $0$ when $s=1$. Thus
\begin{eqnarray}
 \vev{h}  \le 0\,
\end{eqnarray}
and $\vev{h}=0$ only if $S=\mathbf{1}$, i.e.\ $C=C_\infty$. This gives the intuitively obvious result that the best fit for data generated with covariance matrix $C_\infty$ is, on average, given by fitting using the same covariance matrix $C_\infty$. Any other covariance matrix $C$ used for fitting, on average, gives a lower likelihood.

If instead of Eq.~(\ref{eq45}) we had used the difference of $\chi^2$,
\begin{eqnarray}
h_\chi &\equiv& \chi^2_\infty - \chi^2,
\label{eq45a}
\end{eqnarray}
then
\begin{eqnarray}
\vev{h_\chi} &=& N-\text{Tr}\, C^{-1} C_\infty = N-\text{Tr}\, S
=\sum_i \left[1-s_i  \right] \,,\nn
\end{eqnarray}
and $\vev{h_\chi}$ could have either sign, since $s_i > 0$, but need not be smaller than $1$. For example, a simple rescaling,
$C = \lambda C_\infty$, with $\lambda \to \infty$ can always make $\chi^2 \to 0$, its minimum poissible value. This option is eliminated for likelihood because of the $\det (2 \pi C)$ term.

Using for $C$ the best-fit topology $\mathcal{M}_1$ with $L/L_0=1.9$, we find the numerical values
\begin{eqnarray}
N &=& 2482\,,\nn
\ln \det (C_\infty ) &=& 1097.8+ \ln \det C_f\,,\nn
\ln \det (C) &=& 1082.73 + \ln \det C_f \,,\nn
\text{Tr}\, C^{-1} C_\infty &=& \text{Tr}\, S = 2516.6\,,
\end{eqnarray}
 so that
\begin{eqnarray}
\vev{h} &=&  -19.5 \,.
\end{eqnarray}
This differs from the value we find of $h=+20.9$ by $\Delta h=h-\vev{h}=40.4$. The probability that $\Delta h$ is a statistical fluctuation can be determined by computing the variance of $h$ using the four-point function
\begin{eqnarray}
\vev{\Delta_i \Delta_j \Delta_k \Delta_l } &=& \left(C_\infty\right)_{ij} \left(C_\infty\right)_{kl}
+\left(C_\infty\right)_{ik} \left(C_\infty\right)_{jl}\nn
&&+\left(C_\infty\right)_{il} \left(C_\infty\right)_{jk}\,,
\end{eqnarray}
to obtain
\begin{eqnarray}
\vev{\left(\Delta h\right)^2} &=&  2N+2 \text{Tr}\, C^{-1} C_\infty C^{-1} C_\infty  -4 \left(\text{Tr}\, C^{-1} C_\infty\right) \nn
 &=&  2\text{Tr}\,  \left(1-C^{-1} C_\infty\right)^2=2\,\text{Tr}\,  \left(1-S\right)^2\,.
\end{eqnarray}
In our case,
\begin{eqnarray}
\text{Tr}\, C^{-1} C_\infty C^{-1} C_\infty &=& 2610.0\,,
\end{eqnarray}
so that
\begin{eqnarray}
\vev{\left(\Delta h\right)^2} &=&  117.6 = (10.8)^2\,.
\end{eqnarray}
Our observed value of $\Delta h=40.4$ is $3.7\,\sigma$ away from the mean, so the probability that a fluctuation gives $h$ larger than or equal to our observed value is $1.1 \times 10^{-4}$, assuming a normal distribution.

While the distribution of the data $\Delta_i$ is Gaussian, the distribution of the likelihood difference $h$ is no longer Gaussian.
We can also compute higher order connected correlation functions of $h$,
\begin{eqnarray}
\vev{\left(\Delta h\right)^r}_c &=& 2^r\ (r-1)!\ \text{Tr}\, \left(1-C^{-1} C_\infty\right)^r\nn
& = & 2^r\ (r-1)!\ \text{Tr}\, \left(1-S\right)^r\,,
\end{eqnarray}
from the generating function
\begin{eqnarray}
\log \vev{e^{\lambda \Delta h}} &=&  -\lambda \text{Tr}\, (1-C^{-1} C_\infty) \nn
&& - \frac{1}{2} \text{Tr}\, \ln\left[ 1-2\lambda (1-C^{-1} C_\infty) \right] \nn
&=& -\lambda \text{Tr}\, (1-S)  - \frac{1}{2} \text{Tr}\, \ln\left[ 1-2\lambda (1-S) \right] \,,\nn
\end{eqnarray}
so that
\begin{eqnarray}
\vev{(\Delta h)^3} &=&  8\ \text{Tr}\, \left(1-C^{-1} C_\infty\right)^3\nn
\vev{(\Delta h)^4}_c &=&  48\ \text{Tr}\, \left(1-C^{-1} C_\infty\right)^4\,,
\end{eqnarray}
where the fourth-order correlation is
\begin{eqnarray}
\vev{(\Delta h)^4} &=&  \vev{(\Delta h)^4}_c + 3 \vev{(\Delta h)^2}^2\,,
\end{eqnarray}
in terms of the connected correlation. 
The mean value $\vev{h}$, and all the connected correlation functions $\vev{(\Delta h)^r}_c$ are of order $N$, the number of data points. Thus the relative correlation $\vev{(\Delta h)^r}/\vev{h}^r$ is of order $N^{1-r}$.

The numerical values for our case are
\begin{eqnarray}
\vev{\left(\Delta h\right)^3} &=& -489.0\,,\nn
\vev{\left(\Delta h\right)^4}_c &=& 4303.7\,.
\end{eqnarray} 
We can get a better estimate of the probability that $h=20.4$ is due to a statistical fluctuation by using these higher order moments. We have fit $\vev{h}$ and $\vev{(\Delta h)^r}$, $r=2,3,4$ to a probability distribution 
\begin{eqnarray}
p(h) &=& p_0\exp \bigl[-(h-h_0)^2 - c_2 (h-h_0)^2\nn
&&-c_3 (h-h_0)^3-c_4 (h-h_0)^4 \bigr]\,,
\end{eqnarray}
and found using this distribution that the probability that $h - \vev{h} \ge 40.4$ is $10^{-6}$, which is smaller than the value obtained earlier using a normal distribution for $h$.

\section{Dipole Contamination}\label{sec:dipole}

The symmetry axis of $\mathcal{M}_1$ points in the direction $(b=37^\circ\pm2^\circ,l=291^\circ\pm2^\circ)$, which is close to the direction of the velocity of the Local Group $(b=30^\circ \pm 2^\circ, l=276^\circ \pm 3^\circ)$~\citep{Lineweaver:1996qw}. The CMB has a large dipole asymmetry of $3.358 \pm 0.001 \pm 0.023 \, \text{mK}$ in the direction
$(b=48.05^\circ \pm 0.11^\circ, l=264.31^\circ \pm 0.2^\circ)$~\citep{Lineweaver:1996qw}. Suppose that the data is contaminated by a dipole contribution that has not been properly subtracted out.\footnote{This possibility was suggested to us by B.~Keating.} Could a residual dipole explain the results we have found?

To study the effect of a residual dipole, assume that the observed pixels are
\begin{eqnarray}
\Delta_i^{\text{obs}} &=& \Delta_i + \mathbf{p \cdot \hat n}_i = \Delta_i + d_i,\qquad d_i=\mathbf{p \cdot \hat n}_i
\end{eqnarray}
where $\Delta_i$ are the true fluctuations given by the distribution Eq.~(\ref{dist}) and $\mathbf{p}$ is the residual dipole contamination in the data.
Then Eq.~(\ref{eq10},\ref{eq10a}) are replaced by
\begin{eqnarray}
-2 \log \mathcal{L} &=& \left(\Delta +d \right)^T C^{-1}_i \left(\Delta +d \right) + \ln \det (2\pi C)\,,\nn
-2 \log \mathcal{L}_i &=& \left(\Delta +d \right)^T C^{-1}_i \left(\Delta +d \right) + \ln \det (2\pi C)\,.\nn
\end{eqnarray}
From these, we find
\begin{eqnarray}
 \vev{h} 
&=& N-\text{Tr}\, C^{-1} C_\infty + \ln \det (C_\infty )- \ln \det (C)\nn
&& + (d^T C_\infty^{-1} d)-(d^T C^{-1} d) \,,\nn
\vev{\Delta h^2} &=&  2 N - 8 d^T C^{-1} d+4  d^T C_\infty^{-1} d + 4 d^T C^{-1} C_\infty C^{-1}d\nn
&& + 2 \text{Tr}\, C^{-1} C_\infty C^{-1} C_\infty -4 \left(\text{Tr}\, C^{-1} C_\infty\right) \,.
\end{eqnarray}
Dipole contamination  produces a systematic shift in $h$ from its value in Eq.~(\ref{eq47}) given by the
$(d^T C_\infty^{-1} d)-(d^T C^{-1} d)$ terms, which can be written as
\begin{eqnarray}
(d^T C_\infty^{-1} d)-(d^T C^{-1} d) &=& p_\alpha p_\beta D_{\alpha \beta}
\end{eqnarray}
in term of the components $p_\alpha = (p_x,p_y,p_z)$ of the dipole. We find
\begin{eqnarray}
D &=& \left[ \begin{array}{ccc}
 -1.22 & -0.004 & -0.114 \\
 -0.004 & -0.0281 & -0.003 \\
 -0.114 & -0.003 & -0.0127
\end{array}\right]\ \text{mK}^{-2}
\end{eqnarray}
The largest eigenvalue is $-1.23\, \text{mK}^{-2}$. To get a shift in $-2 \ln \mathcal{L}$ of $20.4$ requires a dipole contamination $\abs{\mathbf{p}}$ of around $4\, \text{mK}^{-2}$. This is larger than the observed dipole, and several hundred times the quoted uncertainty in the CMB dipole~\citep{Lineweaver:1996qw}, and is excluded.

\section{Conclusions}\label{conclusions}

We have analyzed the possibility that the universe has compact topologies $\mathcal{M}_0=\mathbb{T}^3$,
$\mathcal{M}_1=\mathbb{T}^2 \times \mathbb{R}^1$ and $\mathcal{M}_2=S^1 \times \mathbb{R}^2$ using modifications of the available CAMB and WMAP 7-year likelihood codes. The maximum likelihood 95\% confidence intervals are $1.7 \le L/L_0 \le 2.1$, $1.8  \le L/L_0 \le  2.0$, $1.2  \le L/L_0 \le 2.1$ for the three cases, respectively. Using the Bayesian analysis discussed earlier, we find that the most probable universe has the compact topology $\mathcal{M}_1$. An infinite universe is compatible with the data at a confidence level of $2 \times 10^{-5}$ (i.e.\  $4.3\,\sigma$). We find no evidence of a preference for the axis-of-evil direction. The improved fit for a finite universe is not due to the lowered prediction for the quadrupole anisotropy; this accounts for only a small fraction of the increase in likelihood.

It would be useful to investigate whether any systematic effects in the WMAP data introduce effects with cubic symmetry that can mimic the effects of a torus topology. The best fit results do not pick out any special orientation for the torus, such as a torus with  symmetry axis perpendicular to the galactic plane, that might lead to systematic effects that lead to a fake signal. Pixelization of the data using the HEALPix grid also should not introduce cubic symmetry terms along an axis not aligned with the galactic pole. The best fit results for $\mathcal{M}_1$ have symmetry axis which is near ($\sim 10^\circ$) the direction of the Local Group velocity.

\acknowledgments

We would like to thank K.~Griest, B.~Keating, and A.~Wolfe for helpful discussions and for comments on the manuscript.
We thank Terrence~Martin and Frank~Wuerthwein for showing us how to use the DOE OSG cluster, the LAMBDA team for making the seven-year WMAP data and the likelihood calculation software available online, and Antony Lewis and Anthony Challinor for making the CAMB software available online.


\bibliography{citations}

\end{document}